# Quantum digital signature scheme

Muhammad Nadeem[1,2*] & Xiaolin Wang[2]
[1]*Department of Basic Sciences, School of Electrical Engineering and Computer Science, National University of Sciences and Technology (NUST), H-12 Islamabad, Pakistan.*
[2]*Spintronic and Electronic Materials Group, Institute for Superconducting & Electronic Materials, Australian Institute for Innovative Materials, University of Wollongong, Wollongong, NSW, 2500, Australia*

Digital signatures are the building blocks of modern communication to prevent masquerading by any party other than recipients, repudiation by signatory and forgery by any individual recipient. Digital signature scheme is said to be standard if the signature (a) is a pattern depending upon the message to be signed, (b) is built upon some information publically known and unique to the signatory, (c) can be stored by all the recipients. While classical methods provide computational security only, quantum mechanics guarantees information-theoretically secure and standard digital signature schemes. However, standard quantum digital signature schemes are based on quantum one-way functions and hence require long term quantum memory for storing quantum signatures, which is not practically feasible yet. We demonstrate here a standard quantum digital signature scheme by replacing quantum one-way functions with multiparty controlled EPR channels. It allows signatory to generate non-locally correlated quantum signatures, instead of multiple copies of a unique quantum state, and assures security against any individual since others have non-locally correlated information.

In general cryptographic setting, secrecy and authentication are the most important security goals. Secrecy assures that encrypted information is unintelligible to eavesdroppers while authentication verifies that information is valid and its originator is genuine. Symmetric cryptography provides assurance of both secrecy (one time pad) and authentication (message authentication codes) if sender and receiver are trusted parties.

However, there can be issues regarding authenticity in case of mutual distrust between sender and receiver. For example, suppose Alice sends a message $m$ to Bob using some authentication technique. Then following disputes can arise: (i) Bob can prepare a different message $m'$ using authentication code and pre-shared secret information and claim that it came from Alice. (ii) Since Bob can change the authenticated message $m$ and generate different message $m'$ appended with authentication code, Alice can deny sending the message as there is no way to resolve this issue of who is cheating.

Hence, among mistrustful sender and receiver, authentication alone is not sufficient to resolve all the issues and something more advanced is required. In general, it is believed that digital signature[1,2] is the solution for this problem where sender appends a code with message that acts his/her signature. Sender's signature assures authenticity and prevents both alteration from receiver or eavesdroppers and denial from the sender. In general, a digital signature scheme must assure that signed message (i) cannot be masquerade by any party other than recipients, (ii) cannot be forged by any individual recipient, (iii) cannot be repudiated by signatory and hence is transferable; if one of the recipients accepts the message as valid and transfers to the others, majority of other recipients must also accept the message as valid.

To fulfill above security requirements, a standard digital signature scheme must be constructed while considering at least following three directions: (a) the signature must be a pattern depending upon the message to be signed. (b) The signature must be built upon some information publically known and unique to the signatory. (c) The scheme must allow all the recipients to store a copy of signature for verification at later stage. The first two requirements assure that the scheme can evade masquerade, repudiation and forgery. In other words, constructing the signature over message being signed with input of some unique information makes it infeasible to create an illegal copy of signature for a signed message or constructing a different message for genuine signature.

For example, Lamport's one-time CDSS[3] based on one-way function[4] works as follows: the signatory secretly chooses $l_0$ and $l_1$ for future single bit message 0 and 1 respectively and publically announces one-way function $f$ and its outcomes $(0, f(l_0))$ and $(1, f(l_1))$ as his/her signature. Since $f$ is one-way function, it is computationally infeasible for anyone else to compute $l_0$ and $l_1$ from $(0, f(l_0))$ and $(1, f(l_1))$ respectively. In the messaging stage, the signatory presents $(m, l_m)$ as his/her signed one-bit message $m$. In the verification phase, any recipient can easily compute $f(l_m)$ and verify that whether it agrees with signatures or not. Since $l_0$ and $l_1$ is something uniquely known only to signatory while function $f$ is publically known, the consistency between $(m, l_m)$ and announced signatures will certify that the message has been sent from the legitimate signatory.

Classical digital signature schemes (CDSS) require public-private key pairs for generating one-way functions[4] where signatures are bit patterns depending on the message being signed with the private key while recipient verifies the signature with signatory's public key. However, CDSS with one-way function (hashing) as main ingredient are only computational secure and can easily be broken with efficient technology; quantum computer[5].

On the other hand, standard quantum digital signature schemes[6,7] (QDSS) based on the laws of quantum mechanics guarantee information-theoretic security for classical message with quantum states being signatures of sender. However, QDSS[6,7] are based on quantum one-way functions (QOWF)[6,8] and hence require quantum memory for storing signatures and later swap tests[7,8] for verification/comparison of signatures in case of dispute. Hence, both of these QDSS[6,7] are practically not feasible with current quantum technologies; either because of swap test or requirement of long term quantum memory.

To overcome the problem of quantum memory, recently an interesting QDSS[9] for classical messages was proposed by using multiport optical techniques similar to those in QDSS[7,10]. The experimental realization of QDSS[9] has also been presented where standard linear optical components and photodetectors are used[11]. Although multiport optical technique avoids long term quantum memory by introducing a new type of quantum measurement, quantum state elimination, however, causes substantial losses when the distance between recipients increases. To overcome loses in QDSS[9,11]; Wallden et al presented QDSS[12] which require neither quantum memory nor a multiport but only commercially available experimental setup similar to those for quantum key distribution. The main idea in QDSS[9-12] is reconsidered recently for multi-bit messages instead of naive iterations of single bit messages[13].

In QDSS[9-12], signatory sends trains of coherent quantum states as his/her signature while recipients stores corresponding classical information, obtained through non-destructive quantum measurements. As a result, it overcomes the need of quantum memory. In the verification stage, signatory sends message and classical information for verification. However, QDSS[9-12] are not standard digital signature schemes unfortunately in their construction. The private classical information stored by signatory is good enough for data authentication but not sufficient for user authentication. Such schemes, without incorporating security requirement (b), cannot assure user authentication while making public decisions. Hence, security against masquerading cannot be guaranteed over public channels, which is a serious drawback as for as standard digital signature scheme is concerned. However, QDSS[9-12] can overcome masquerading, forgery as well as repudiation under very strict conditions; all quantum/classical channels are authenticated. It should not be the case with standard digital signature scheme.

In short, QOWF guarantee information-theoretic security against masquerading as well as forgery and repudiation but make quantum memory an essential component of standard QDSS. Hence, in order to avoid the requirement of quantum memory for practically feasible QDSS, a scheme must be devised that fulfills all the standard security requirements (a), (b) and (c) but without QOWF.

To circumvent this problem with standard QDSS in particular and asymmetric quantum cryptography in general, we demonstrate here a practically efficient and information-theoretic standard QDSS based on multiparty controlled EPR channel[14,15]. Over multiparty controlled EPR channels, generated by entanglement swapping[16,17] and teleportation[17], signatory Alice and receivers Bob and Charlie all have some pieces of control. It allows Alice to generate non-locally correlated quantum signatures for Bob and Charlie; signature states received by Bob and Charlie are not the multiple copies of unique quantum state but are different states non-locally correlated with each other. Such non-locally correlated signatures assure security against masquerading, forgery as well as repudiation and hence transferability is guaranteed. In general, multiparty controlled channel does not allow masquerading where signatory and receivers have non-locally correlated information in a unique fashion.

Finally, all the existing QDSS are inspired by CDSS based on public-private key systems where signatures are generated by message and private key of signatory while recipient verifies the signature with signatory's public key. However, our proposed QDSS is similar to digital signature standard (DSS) where a pair of signatures is generated from the message being signed while the recipients authenticate the message by comparing the signatures. However, our proposed QDSS is more efficient than DSS and different in construction, there is no classical counterpart of quantum non-local correlations. Moreover, in the proposed QDSS, signatory does not require to prepare public-private key pair and distribute public key before signing a message. Instead, he/she teleports the message over multiparty controlled EPR channels which directly results in a pair of non-locally correlated signatures.

**Multiparty controlled EPR channel**

Suppose Alice and Bob share a publically known EPR channel $|\Theta_{ab}\rangle \in \{|\Phi^+\rangle, |\Psi^+\rangle, |\Phi^-\rangle, |\Psi^-\rangle\}$ between them where $|\Phi^\pm\rangle = (|00\rangle \pm |11\rangle)/\sqrt{2}$ and $|\Psi^\pm\rangle = (|01\rangle \pm |10\rangle)/\sqrt{2}$. In standard teleportation[17], Alice performs Bell state measurement (BSM)[18] on quantum state $|\psi_a\rangle$ and her half of EPR pair such that she gets classical 2-bits $aa' \in \{00,01,10,11\}$ while Bob's half becomes $|\psi_b\rangle = \sigma |\psi_a\rangle$ where $\sigma \in \{I, \sigma_x, \sigma_z, \sigma_z\sigma_x\}$ is Pauli operator. If Alice knows state $|\psi_a\rangle$, she also knows the state $|\psi_b\rangle = \sigma |\psi_a\rangle$ on Bob's side since Pauli encoding $\sigma$ is correlated with her classical BSM result $aa'$.

In other words, EPR channel $|\Theta_{ab}\rangle$ is fully controlled by only one party, sender Alice here. Standard quantum teleportation has fascinating applications in quantum cryptography, information theoretic secrecy say, if sender and receiver are trusted. However, in case of mutual distrust between sender and receiver, there can be following issues: (i) the sender can repudiate by altering both state $|\psi_a\rangle$ and her BSM result $aa'$; especially when delay between teleportation and announcement of classical information $aa'$ is required by the cryptographic task. (ii) Even if Alice make classical information $aa'$ public soon after teleportation occurs; still there is no way to overcome alteration from Bob and denial from Alice.

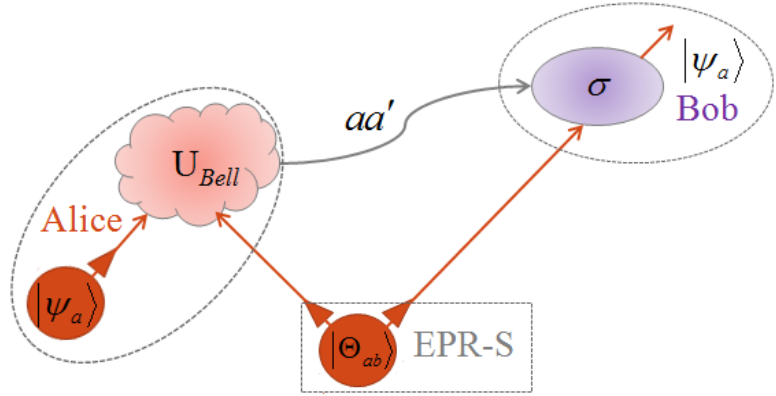

**Figure 1:** Standard Teleportation: Alice and Bob share EPR pair $|\Theta_{ab}\rangle$ generated from some EPR source and Alice performs joint measurement on state $|\psi_a\rangle$ and her half of EPR pair such that she gets classical 2-bits $aa'$ while Bob's half becomes $|\psi_b\rangle = \sigma|\psi_a\rangle$. If Alice sends her measurement result $aa'$ to Bob, he can get $|\psi_a\rangle$ by applying corresponding Pauli operator on his half.

By performing entanglement swapping and teleportation together, a multiparty controlled EPR channel can be established where all three parties Alice, Bob, and Charlie have shares and neither party alone can control this channel[14]. Suppose a Bell state $|\Theta_{ac}\rangle$ is shared between Alice and Charlie while Bell state $|\Theta_{bc'}\rangle$ is shared between Bob and Charlie. If Charlie performs BSM on his halves, he gets two classical bits $cc' \in \{00,01,10,11\}$ while one of the four EPR channel $|\Theta_{ab}\rangle = \sigma_c|\Phi_{ab}^+\rangle$ swaps between Alice and Bob. Now if Alice teleports a quantum state $|\psi_a\rangle$ to Bob over channel $|\Theta_{ab}\rangle$, Alice gets two classical bits $aa' \in \{00,01,10,11\}$ while Bob's half becomes one of the four possibilities $|\psi_b\rangle = \sigma_a\sigma_c|\psi_a\rangle$ where Pauli encoding $\sigma_a\sigma_c$ is correlated with both $aa'$ and $cc'$.

Here, the control of the EPR channel $|\Theta_{ab}\rangle$ is shared between all three parties Alice, Bob and Charlie. Charlie keeps classical information $cc'$, and hence knows the exact identity of channel $|\Theta_{ab}\rangle$. Alice possesses classical information $aa'$ and message state $|\psi_a\rangle$ but don't know the encrypted message $|\psi_b\rangle = \sigma_a\sigma_c|\psi_a\rangle$ kept by Bob. Hence, Alice cannot simulate (repudiate) her alterations in state $|\psi_a\rangle$ and BSM result $aa'$ with $|\psi_b\rangle = \sigma_a\sigma_c|\psi_a\rangle$. Similarly, Bob keeps only state $|\psi_b\rangle = \sigma_a\sigma_c|\psi_a\rangle$ but remains unknown to other shares $aa'$ and $cc'$ and hence Pauli encoding $\sigma_a\sigma_c$ unless both Alice and Charlie reveal their secrets.

When both Alice and Bob present their shares $|\psi_a\rangle$, $aa'$ and $|\psi_b\rangle = \sigma_a \sigma_c |\psi_a\rangle$ respectively to Charlie, he can deduce whether $|\psi_b\rangle = \sigma_a \sigma_c |\psi_a\rangle$ is consistent with non-locally correlated shares $aa'$ and $cc'$ or not. Such a multiparty controlled EPR channel guarantees security against masquerading, forgery from Bob and repudiation from Alice and hence allows unconditional transferability as demonstrated in next section where take $|\Theta_{ac}\rangle = |\Theta_{bc'}\rangle = |\Phi^+\rangle$. Hence,

$$|\psi_b\rangle = \sigma_z^{a \oplus c} \sigma_x^{a' \oplus c'} |\psi_a\rangle \tag{1}$$

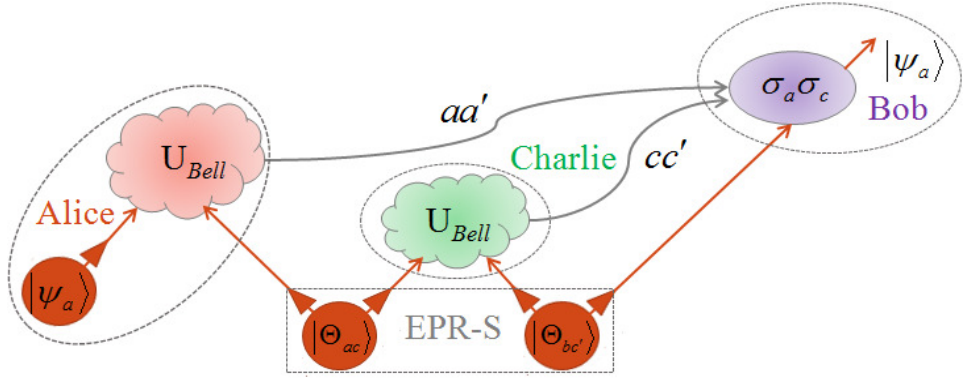

**Figure 2:** Setup for multiparty controlled EPR channel where Alice and Charlie share EPR pair $|\Theta_{ac}\rangle$ while Bob and Charlie share $|\Theta_{bc'}\rangle$. Charlie performs BSM on his entangled halves and Alice performs BSM on the state $|\psi_a\rangle$ and her half of EPR pair such that Bob's half becomes $|\psi_b\rangle = \sigma_a \sigma_c |\psi_a\rangle$. Bob can only apply exact Pauli operator on his half and get the state $|\psi_a\rangle$ if and only if he receives BSM results from both Alice and Charlie.

**Quantum digital signature scheme**
We demonstrate here a standard quantum signature scheme between sender Alice and two receivers Bob and Charlie. In the next section, we showed that the proposed QDSS guarantees information theoretic security against masquerading as well as repudiation and forgery.

**Multiparty controlled EPR channels:** Suppose publically known $n$ EPR pairs $|\Phi_{ac}^+\rangle^{\otimes n}$ are shared between Alice and Charlie while $n$ EPR pairs $|\Phi_{bc'}^+\rangle^{\otimes n}$ are shared between Bob and Charlie respectively. Charlie performs Bell state measurement on respective halves, stores BSM result $c = \otimes_{i=1}^{n} c_i c_i'$ as his private key, and sends $c_p = \otimes_{i=1}^{n} c_i \oplus c_i'$ to Bob securely. As a result, $n$ EPR pairs $|\Theta_{ab}\rangle^{\otimes n}$ swaps between Alice and Bob whose exact identity is known only to Charlie.

**Non-locally correlated Signature distribution:** Alice prepares a quantum state $|m_a\rangle = |m_1\rangle |m_2\rangle .... |m_n\rangle$ corresponding to her classical message $m_a = m_1 m_2 .... m_n$ where $m_i \in \{0,1\}$ and generates $|\psi_a\rangle = \otimes_{i=1}^{n} U_i |m_i\rangle$ by applying operator $U$ qubitwise. The operator $U$ is publically known and acts as a unitary transformation from computational basis $\{0,1\}$ to $\{\delta_0, \delta_1\}$ basis;

$\delta_0 = U|0\rangle = (|0\rangle + i|1\rangle)/2$ and $\delta_1 = U|1\rangle = (|0\rangle - i|1\rangle)/2$. Alice generates and distributes her non-locally correlated signatures as follows:

(i) Alice teleports state $|\psi_a\rangle$ to Bob over EPR channel $|\Theta_{ab}\rangle^{\otimes n}$ and stores BSM result $a = \overset{n}{\underset{i=1}{\otimes}} a_i a'_i$ and $a_p = \overset{n}{\underset{i=1}{\otimes}}(a_i \oplus a'_i)$ as her private key pair $(a, a_p)$. As a result, entangled halves in possession of Bob become either $|\psi_b\rangle = \overset{n}{\underset{i=1}{\otimes}} \sigma_z^{a_i \oplus c_i} \sigma_x^{a'_i \oplus c'_i}|\psi_{a_i}\rangle$. Bob measures $|\psi_b\rangle$ in $\{\delta_0, \delta_1\}$ basis and stores $n$-bit strings $S_b$ as Alice's signature

(ii) Alice generates her global signature,

$$|\psi_a^G\rangle = \overset{n}{\underset{i=1}{\otimes}} \sigma_z^{a_i} \sigma_x^{a'_i}|\psi_{a_i}\rangle \qquad (2)$$

measures in $\{\delta_0, \delta_1\}$ basis and announces the outcome $S_a^G$.

**Verification:** Alice sends her secret pair $\{m'_a, a'_p\}$, possibly altered, to Bob. Bob calculates $S_a'^G$ from equation (2) and verifies whether message is genuine or repudiated by comparing Alice's signature $S_b$ and $S_a^G$:

$$v_1; S_{a_i}'^G = S_{a_i}^G \qquad (3)$$

$$v_2; \begin{cases} S_{a_i}^G = S_{b_i}; c_{p_i} = 0 \\ S_{a_i}^G \neq S_{b_i}; c_{p_i} = 1 \end{cases} \qquad (4)$$

If Bob authenticates the message with very high probability, he forwards the triplet $\{m'_a, a'_p, S'_b\}$, possibly altered, to Charlie. Charlie concludes whether forgery or repudiation has occurred or not as follows: Charlie calculates $S_b$ from

$$\sigma_x^{c'_i} \sigma_z^{c_i}|\psi_{a_i}^G\rangle = |\psi_{b_i}\rangle \qquad (5)$$

and verifies

$$v_3; S_{b_i} = S'_{b_i} \qquad (6)$$

If $v_3$ is satisfied with very high probability, he accepts the message genuine and secure against forgery from Bob and repudiation from Alice by verifying functions $v_1$ and $v_2$ (3,4). (*Warning: To authenticate no forgery, Charlie should not rely on Bob's triplet $\{m'_a, a'_p, S'_b\}$ only to verify both functions $v_1$ and $v_2$.*)

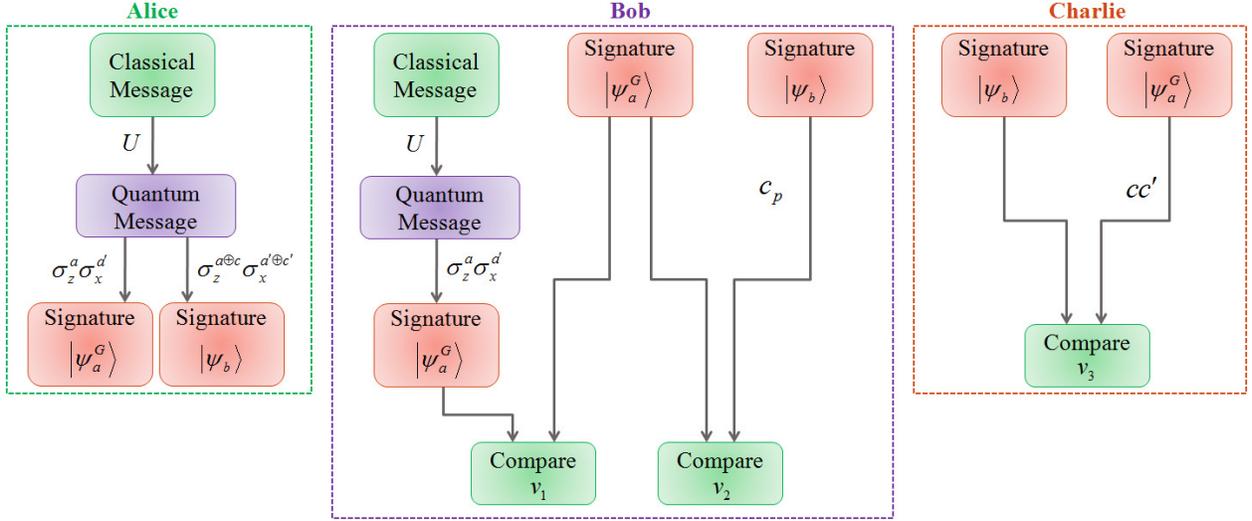

**Figure 3:** Schematics of proposed quantum digital signature scheme.

**Security Analysis:**
In our proposed standard QDSS, information-theoretic security against forgery, repudiation and masquerading comes from the fundamental rules of quantum mechanics instead of computational hardness. Instead of distributing multiple copies of a unique quantum signature, non-locally correlated signatures guarantees no forgery, non-repudiation and hence transferability. Finally, the pre-shared EPR states and non-locally correlated signatures do not allow masquerades to impersonate by sending an illegal copy of signature for a signed message or constructing a different message for genuine signature.

**Theorem:** *Suppose Alice and Charlie share publically known EPR pairs $|\Phi^+_{ac}\rangle^{\otimes n}$ while Bob and Charlie share publically known EPR pairs $|\Phi^+_{bc'}\rangle^{\otimes n}$. Charlie performs BSM on his entangled halves and Alice performs joint measurement on the state $|\psi_a\rangle = \overset{n}{\underset{i=1}{\otimes}} U_i |m_i\rangle \in \{\delta_0, \delta_1\}^{\otimes n}$ and her halves of EPR paisr such that Charlie keeps 2n-bits $c = \overset{n}{\underset{i=1}{\otimes}} c_i c'_i$, Alice gets 2n-bits $a = \overset{n}{\underset{i=1}{\otimes}} a_i a'_i$ while Bob's half becomes $|\psi_b\rangle = \overset{n}{\underset{i=1}{\otimes}} \sigma_z^{a_i \oplus c_i} \sigma_x^{a'_i \oplus c'_i} |\psi_{a_i}\rangle$. Bob stores n-bit string $S_b$ as classical counterpart of Alice signature $|\psi_b\rangle$ while Alice announces classical counterpart $S_a^G$ of state $|\psi_a^G\rangle = \overset{n}{\underset{i=1}{\otimes}} \sigma_z^{a_i} \sigma_x^{a'_i} |\psi_{a_i}\rangle$ as her global signature publically.*

*(i) **Non-repudiation:** If Alice don't know $c_p = \overset{n}{\underset{i=1}{\otimes}} (c_i \oplus c'_i)$, she cannot change pair $\{m_a, a_p\}$ such that both verification functions $v_1$ and $v_2$ are satisfied with very high probability.*

*(ii) **Transferability:** If Bob accepts the message genuine and transfers, Charlie will also accept the message valid with very high probability.*

*(iii) No forgery:* Even if Bob knows both $a_p = \bigotimes_{i=1}^{n}(a_i \oplus a'_i)$ and $c_p = \bigotimes_{i=1}^{n}(c_i \oplus c'_i)$, he cannot change triplet $\{m_a, a_p, S_b\}$ such that all three verification functions $v_1$, $v_2$, and $v_3$ are satisfied with very high probability.

*(iv) No masquerading:* Any third party other than recipients Bob and Charlie, cannot impersonate Alice.

**Proof:** Alice's signatures $|\psi_b\rangle$ and $|\psi_a^G\rangle$ are not the copies of unique quantum state but are two different quantum states non-locally correlated with each other. From equation (1,2)

$$|\psi_{a_i}^G\rangle = \sigma_z^{c_i}\sigma_x^{c'_i}|\psi_{b_i}\rangle \tag{7}$$

Above relation between Alice's signatures is independent of her BSM result $aa'$. Moreover, basis $\delta_0$ and $\delta_1$ are eigenbasis of Pauli operator $\sigma_z\sigma_x$ with eigenvalues $i$ and $-i$ respectively, $\sigma_z$ acts as not gate while $\sigma_x$ acts as not plus phase flip gate in $\{\delta_0, \delta_1\}$ basis; $\sigma_x|\delta_0\rangle = i|\delta_1\rangle$ and $\sigma_x|\delta_1\rangle = -i|\delta_0\rangle$. That is,

$$\sigma_z\sigma_x|\delta_i\rangle = (-1)^i i|\delta_i\rangle \tag{9}$$

$$\sigma_z|\delta_i\rangle = |\delta_{i\oplus 1}\rangle \tag{10}$$

$$\sigma_x|\delta_i\rangle = (-1)^i i|\delta_{i\oplus 1}\rangle \tag{11}$$

Since Alice do not know the Charlie's private key $(c, c_p)$, he cannot extract the Pauli encoding in equation (7).

- If $c_{p_i} = 0$, then $|\psi_{a_i}^G\rangle = |\psi_{b_i}\rangle \Leftrightarrow S_{a_i}^G = S_{b_i}$.
- If $c_p = 1$, then $|\psi_{a_i}^G\rangle \neq |\psi_{b_i}\rangle \Leftrightarrow S_{a_i}^G \neq S_{b_i}$.

*(i) Non-repudiation:* In the proposed QDSS Alice has very little resources to repudiate the message successfully: Charlie's private key $(c, c_p)$ and hence Bob's share $S_b$ are unknown to Alice. Moreover, Alice's signatures $S_b$ and $S_a^G$ are correlated in a fashion such that unitary transformation between them are independent of her BSM result $aa'$ (7). Hence, after distributing her global signature $S_a^G$, she cannot repudiate the message or create a dispute between Bob and Charlie by altering her pair $\{m_a, a_p\}$ such that both verification functions $v_1$ and $v_2$ are satisfied with very high probability. There will be following two possibilities:

- If Alice sends genuine pair $\{m_a, a_p\}$, both verification functions $v_1$ and $v_2$ will be satisfied with very high probability.
- If Alice sends altered pair $\{m'_a, a'_p\}$, verification functions $v_1$ can be satisfied but verification functions $v_2$ will FAIL with very high probability.

$$p_{repudiation} \leq \left(\frac{1}{2}\right)^n \tag{8}$$

*(ii) Transferability:* Proposed QDSS is secure against Alice's attempts to repudiate the message and hence fulfils the condition of transferability: if Bob accepts the message valid and transferable, then Charlie will also accept the message valid with very high probability. That is, authentication of Bob means verification function $v_2$ is satisfied with very high probability. Hence verification function $v_3$ should also be verified by Charlie if Bob is not forging.

*(iii) No forgery:* The most fascinating equalities in the proposed QDSS are (6) and (7); Charlie does not require $\{m_a, a_p\}$ either from Alice or from Bob to authenticate verification function $v_3$. Non-local correlations generated through multiparty controlled EPR channel allow Charlie to extract Bob's share $S_b$ by using publically announced Alice's global signature $S_a^G$ and his own private key $(c, c_p)$. As a result, the verification function $v_3$ bounds Bob from forging triplet $\{m_a, a_p, S_b\}$. If Bob does, all three verification functions $v_1$, $v_2$, and $v_3$ will NOT be satisfied.

*(iv) No masquerading:* Since EPR pairs $|\Phi_{ac}^+\rangle^{\otimes n}$ and $|\Phi_{bc'}^+\rangle^{\otimes n}$ are publically known, multiparty controlled EPR channels does not allow masquerading where signatory and receivers have non-locally correlated information in a unique fashion.

**QOWF Vs multiparty controlled EPR channel**
Quantum one-way functions can be obtained by defining a map $s \to |\psi_s\rangle$ with classical *r*-bit string *s* as input and *t*-qubit quantum string $|\psi_s\rangle$ as output[5,7]. By setting *r* exponentially longer than *t*, generating nearly orthogonal states $|\psi_s\rangle$ and $|\psi_{s'}\rangle$ for $s \neq s'$, it becomes impossible to invert the map $s \to |\psi_s\rangle$. That is, $|\psi_s\rangle$ is easy to generate by knowing *s* but hard to get *s* by knowing $|\psi_s\rangle$.

For establishing such an information-theoretic QOWF, Holevo's theorem[19] is the fundamental principle which shows that there is one-to-one correspondence between quantum and classical processing units; measurement on a single qubit can give at most single classical bit of information. Hence *t*-qubit quantum string $|\psi_s\rangle$ can give maximum of *t*-bits but not classical string $r >> t$. As a result, QOWF guarantees information-theoretic security against masquerading, forgery as well as repudiation but make quantum memory an essential component of standard QDSS.

On the other hand, multiparty controlled EPR channel achieves same goals as QOWF guarantees but without the requirement of quantum memory. Each qubit state $|\psi_{a_i}\rangle$ is associated with unique classical bit *m*. Hence mapping $m \to |\psi_{a_i}\rangle$ is two-way but teleporting such states $|\psi_{a_i}\rangle$ over EPR channel controlled by multiparty allows generating two different quantum states that are non-locally correlated with each other where all three parties have some shares. Hence it becomes infeasible for any individual, signatory as well as recipient, to alter the state $|\psi_{a_i}\rangle$ in a deterministic way.

**Discussion:**
We discussed that all the existing standard quantum digital signature schemes are based on quantum one-way functions. As a result, laws of quantum mechanics guarantee information-theoretic security for classical message with quantum states as being signatures of sender.

However, the price of using quantum one-way functions is requirement of quantum memory for storing signatures and later swap tests for verification or comparison of signatures in case of dispute.

We also discussed that there exist some quantum digital signature schemes without quantum memory or quantum one-way functions that are easy to demonstrate practically but such schemes are not standard; signatory keeps private classical information but it is not unique to him/her while making public decisions, enough for data authentication but not sufficient for user authentication. Hence, security against masquerading cannot be guaranteed over public channels, which is a serious drawback as for as standard digital signature scheme is concerned.

We then proposed an information-theoretic quantum digital signature scheme replacing quantum one-way functions with multiparty controlled EPR channels but without compromising standard security requirements for any practically feasible scheme. We showed that such EPR channels controlled by multiparty guarantees information-theoretic security by allowing two different quantum states as signatures of sender that are non-locally correlated with each other where all three parties have some shares. Hence it becomes infeasible for any individual, signatory as well as recipient, to alter the message or signature in a deterministic way. More importantly, signatures are not the copies of unique quantum state, as all previously existing QDSS have, but are two different but non-locally correlated states.

Instead of using many-to-one mapping as QOWF does, we use one-to-one mapping of classical bits and quantum bits. This one-to-one mapping along with multiparty controlled EPR channels gives information-theoretic security and removes the requirement of quantum memory. Every recipient can store classical message in his classical memory and regenerate corresponding quantum message with the help of global public key whenever required.

Moreover, unlike all the existing QDSS, proposed QDSS does not bound signatory to prepare public-private key pair and distribute public key before signing a message. Instead, he/she teleports the message over multiparty controlled EPR channels which directly results in a pair of non-locally correlated signatures and allow recipient to compare and verify message. This technique of generating non-locally correlated pair of states would also allow asymmetric quantum cryptography to become practically feasible and secure in general; public-private key pair generated through QOWF does not allow distributing indefinite copies of public key as classical asymmetric cryptography doe's.

**Acknowledgments**
We acknowledge the support from the Australian Research Council (ARC) through an ARC Discovery Project (DP130102956) and an ARC Professorial Future Fellowship project (FT130100778).